\documentclass[]{article}
\usepackage{amssymb}
\def\ba{\mathbf{a}}   
\def\bb{\mathbf{b}}   
\def\bc{\mathbf{c}}   
\def\be{\mathbf{e}}   
\def\bm{\mathbf{m}}   
\def\bn{\mathbf{n}}   
\def\bp{\mathbf{p}}   
\def\br{\mathbf{r}}   
\def\bs{\mathbf{s}}   
\def\bt{\mathbf{t}}   
\def\bv{\mathbf{v}}   
\def\bx{\mathbf{x}}   
\def\bB{\mathbf{B}}   
\def\bT{\mathbf{T}}   

\def\C{\mathbb{C}} 
\def\G{\mathbb{G}} 

\def\R{\mathbb{R}}  

\def\no{\noindent}

\def\beq{\begin{equation}}
\def\eeq{\end{equation}}
\def\con{\overline}

\def\w{\wedge}

\usepackage{helvet}
\usepackage{courier}
\usepackage{graphicx}   
\usepackage{makeidx}
\usepackage{multicol} 
\usepackage{mathptmx} 
\def\bpm{\begin{pmatrix}}
\def\epm{\end{pmatrix}}
\newtheorem{uncertainty}{Uncertainty Principle}[section]
\makeindex            
\begin{document}
\title{Part I: Vector Analysis of Spinors}
\author{Garret Sobczyk
\\ Universidad de las Am\'ericas-Puebla
 \\ Departamento de F\'isico-Matem\'aticas
\\72820 Puebla, Pue., M\'exico
\\ http://www.garretstar.com}
\maketitle
\begin{abstract}Part I:  The geometric algebra $\G_3$ of space is derived by extending the real number system 
to include three mutually anticommuting square roots of $+ 1$. The resulting geometric
algebra is isomorphic to the algebra of complex $2\times 2$ matrices, also known as the Pauli algebra.
The so-called spinor algebra of $\C_2$, the language of the quantum mechanics, is formulated in terms of the
idempotents and nilpotents of the geometric algebra $\G_3$, including its beautiful representation on the Riemann sphere,
and a new proof of the Heisenberg uncertainty principle. In Part II: {\it Spacetime Algebra of Dirac Spinors}, the ideas are
generalized to apply to $4$-component Dirac spinors, and their geometric interpretation in spacetime.

\smallskip

\smallskip

\no {\em Keywords:} bra-ket formalism, geometric algebra, Schr\"odinger-Pauli equation, spinor, 
spacetime algebra, Minkowski spacetime, Riemann sphere.

\smallskip

\no {\em AMS Subject Classication 2010:} 15A66, 81P16 

\end{abstract}

\section{Introduction}
Three dimensional Gibbs-Heaviside vector analysis was developed early in the 20th Century, before
the development of relativity and quantum mechanics. What is still not widely appreciated is that
the Gibbs-Heaviside formalism can be fortified into a far more powerful geometric algebra which
serves the much more sophisticated needs of relativity theory and quantum mechanics. The associative
Geometric algebra is viewed here as the natural completion of the real number system to include the concept
of direction. 

We assume that the real numbers can be always be extended to include new {\it anticommuting} square roots of
$+1$ and $-1$. The new square roots of $+1$ represent orthogonormal Euclidean vectors along the
$xyz$-coordinate axes, whereas new square roots of $-1$ represent orthogonormal pseudo-Euclidean vectors
along the coordinate axes of more general pseudo-Euclidean spacetimes. We shall primarily be
interested in the geometric algebra $\G_3$ of the ordinary $3$-dimensional space $\R^3$ of experience,
but the interested reader may pursue how this geometric algebra can be {\it factored} into the spacetime
algebra $\G_{1,3}$ of the pseudo-Riemannian space $\R^{1,3}$ of $4$-dimensional Minkowski spacetime, \cite[Ch.11]{SNF}. 

One of the most important concepts in quantum mechanics is the concept of spin, and the
treatment of spin has led to many important mathematical developments, starting with the
Pauli and Dirac matrices in the early development of quantum mechanics, to the development of the
differential forms, geometric algebras, and other more specialized formalisms, such as the twistor formalism 
of Roger Penrose \cite[Ch.33]{Pen04}. We show here how the geometric algebra $\G_3$ of $3$-dimensional Euclidean space $\R^3$
 has all the algebraic tools
necessary to give a clear geometrical picture of the relationship between a classical $2$-component {\it spinor}
in the complex plane, and a point on the Riemann sphere obtained by stereographic projection
from the South Pole.

So let's get started.

\section{Geometric algebra of space}

We extend the real number system $\R$ to include three new {\it anticommuting} square roots $\be_1, \be_2$, $\be_3$ of $+1$,
which we identify as unit vectors along the $x$- $y$- and $z$-axis of Euclidean space $\R^3$. Thus,
\[  \be_1^2=\be_2^2=\be_3^2=1, \quad {\rm and} \quad \be_{jk}:=\be_j\be_k=-\be_k \be_j=- \be_{kj}\]
for $1 \le j < k \le 3$.
We assume that the associative and distributive laws of multiplication of real numbers remain valid
in our geometrically extended number system, and give the new quantities $I:= \be_{23}$, $J:= \be_{13}$
and $K:=\be_{12}$ the geometric interpretation of {\it directed plane segments}, or {\it bivectors}, parallel
to the respective $yz$-, $xz$- and $xy$-planes. 
Every unit bivector 
is the generator of rotations in the vector plane of that
bivector, and this property generalizes to bivectors of the $n$-dimensional Euclidean space $\R^n$.  
We leave it for the reader to check that $I,J,K$ satisfy
exactly the same rules as Hamilton's famous quaternions, but now endowed with the geometric interpretation
of oriented bivectors, rather than Hamilton's original interpretation of these quantities as the unit vectors
$\be_1, \be_2, \be_3$.

Whereas the unit bivectors $I,J,K$ satisfy $I^2=J^2=K^2 = -1$, the new quantity
$i:= \be_{123}:=\be_1 \be_2 \be_3$ is a unit {\it trivector}, or {\it directed
volume} element. We easily calculate, with the help of the associative and anticommutitive properties, 
\[ i^2 = (\be_1 \be_2 \be_3)(\be_1 \be_2 \be_3) =\be_1^2 \be_2 \be_3 \be_2 \be_3 =- \be_1^2 \be_2^2 \be_3^2
 = -1,\]
so the unit trivector $i$ has same square minus one as do the bivectors $I,J,K$.

The important Euler identity
 \[ e^{i \theta}=\cos \theta + i \sin \theta  \]
 for $\theta \in \R$, depends only upon
 the algebraic property that $i^2=-1$, and so is equally
 valid for the unit bivectors $I,J,K$.
 For the unit vectors $\be_k$, we have the {\it hyperbolic Euler identities} 
\[ e^{\be_k \phi}=\cosh \phi + \be_k \sinh \phi  \]
for $\phi \in \R$ and $k=1,2,3$.
All of these identities are special cases of the general algebraic definition of the exponential function
\[  e^X\equiv \sum_{n=0}^\infty \frac{X^n}{n!}= \cosh X + \sinh X, \]
\cite[Chp. 2]{SNF} and \cite{S1}.

\begin{figure}
\begin{center}
\includegraphics[scale=.25]{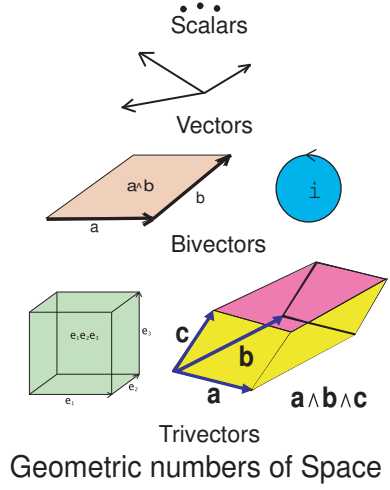}
\caption{Geometric numbers of space}
\label{picbasis}
\end{center}
\end{figure}

The {\it standard basis} of the $2^3=8$ dimensional geometric algebra $\G_3$,
 with respect to the {\it coordinate frame} $\{\be_1,\be_2,\be_3\}$ of
the Euclidean space $\R^3$, is
\beq \G_3:=\G(\R^3)=span\{1,\be_1, \be_2, \be_3,\be_{12},\be_{13},\be_{23}, \be_{123} \} . \label{geospace} \eeq
The geometric numbers of $3$-dimensional space are pictured in Figure \ref{picbasis}.

 Alternatively, we can obtain the geometric algebra $\G_3$ by its representation as a
$2\times 2$ matrix algebra $Mat_\C(2)$ over the complex numbers. 
We define the mutually anihiliating idempotents $u_\pm := \frac{1}{2}(1 \pm \be_3)$,
and note the fundamental relationships 
\beq u_+u_-=0,\quad u_+ + u_- =1, \quad u_+ - u_- = \be_3, \quad {\rm and} \quad  \be_1 u_+ = u_- \be_1. \label{funrelations} \eeq
 Since the unit trivector $i$ commutes
with all the elements (is in the center) of $\G_3$, 
the {\it spectral basis} of $\G_3$, over the {\it fomally} complex numbers $\C=span_\R\{1,i\}$, is specified by
\beq \G_3 = span\{\pmatrix{1 \cr \be_1}u_+ \pmatrix{1 & \be_1}\}= span\{ \pmatrix{u_+ & \be_1 u_- \cr
                                            \be_1 u_+ & u_-} \}. \label{specbasis3} \eeq
                                            
 The relationship between the {\it standard basis} (\ref{geospace}) and the spectral basis (\ref{specbasis3}) is
directly expressed by 
\beq \pmatrix{1 \cr \be_1 \cr \be_2 \cr \be_3}=\pmatrix{1 & 0 & 0 & 1 \cr 0 & 1 & 1 & 0 \cr 0 & i & -i & 0 \cr
         1 & 0 & 0 & -1}\pmatrix{u_+ \cr \be_1u_+ \cr \be_1 u_- \cr u_-}. \label{tablebasis3}  \eeq
For example, using the relationships (\ref{funrelations}), the spectral basis (\ref{specbasis3}),
 and the fact that
\[\be_2=\be_{123} \be_1 \be_3 =i\,\be_1(u_+- u_-) ,\]
the famous {\it Pauli matrices} of the coordinate frame $\{\be_1,\be_2,\be_3\}$
 are simply obtained,
 getting
\beq [\be_1] := \pmatrix{0 & 1 \cr 1 & 0}, \ [\be_2]:= \pmatrix{0 & -i  \cr i & 0}, \ [\be_3]:= -i[\be_1][\be_2]=
              \pmatrix{1 & 0 \cr 0 & -1}. \label{paulimatrices1} \eeq  

       Indeed, the matrix representation $[g]\in Mat_\C(2)$ of any geometric number $g \in \G_3$ is
\beq  g= \pmatrix{1 & \be_1}u_+[g]\pmatrix{1 \cr \be_1} . \label{matrixrep} \eeq 
For example, the unit vector $\be_2$ of the Pauli matrix $[\be_2]$ in (\ref{paulimatrices1}),
  is specified by
\[     \be_2=  \pmatrix{1 & \be_1}u_+ \pmatrix{0 & -i  \cr i & 0}\pmatrix{1 \cr \be_1}  
= i \be_1 u_+ - i\, \be_1 u_- =i\,\be_1(u_+- u_-)\]
in agreement with (\ref{tablebasis3}). The proof of the isomorphism of the complex matrix
algebra $Mat_\C(2)$ and the geometric algebra $\G_3$ is left to the reader. For a further discussion, see \cite[p.79]{SNF}.                             

Any geometric number $g\in \G_3$ can be written in the form 
\beq g=\sum_{k=0}^3 \alpha_k \be_k \label{coorformg} \eeq
 where $\be_0 :=1$, $\alpha_k = a_k+i\, b_k$ for $a_k, b_k \in \R$, $i=\be_{123}$,
  and where $0 \le k \le 3$, giving $2^3=8$ degrees of freedom. 
 The conjugation, known as the {\it reverse} $g^\dagger$ of
the geometric number $g$, is defined by reversing the orders of all the products of the
 vectors that make up $g$, giving
 \beq g^\dagger := \sum_{k=0}^3 \con\alpha_k \be_k . \label{conjuofrev} \eeq 
 In particular, writing $g=s + \bv + \bB+ \bT$, the sum of a real number $s\in \G_3^0$, a vector $\bv\in \G_3^1$,
  a bivector $\bB\in \G_3^2$
 and a trivector $\bT\in \G_3^3$, $g^\dagger = s + \bv - \bB - \bT$.
 
 Two other conjugations are widely used in geometric algebra. The {\it grade inversion} is
  obtained by replacing each vector in a product by
 its negative. It corresponds to an inversion in the origin, otherwise known as a {\it parity inversion}. For the
 geometric number $g$ given in (\ref{coorformg}), the grade inversion is 
 \beq g^- :=\con \alpha_0 - \sum_{k=1}^3 \con \alpha_k \be_k.   \label{coninversion} \eeq
 When $g$ is written as $g=s + \bv + \bB+ \bT$, the grade inversion $g^-=s-\bv +\bB-\bT$.
 The {\it Clifford conjugation} $g^*$ of the geometric number $g\in \G_3$, defined by
 \beq g^* := (g^-)^\dagger = \alpha_0-\sum_{k=1}^3 \alpha_k \be_k= s-\bv -\bB +\bT,      \label{clifforcon} \eeq
 is just the inversion of $g$ followed by the reversion.

All other products in the geometric algebra are defined in terms of the geometric product.
For example, given vectors $\ba, \bb \in \G_3^1\equiv\R^3$,
\beq \ba \bb = \ba \cdot \bb + \ba \w \bb \in \G_3^{0+2} ,      \label{gaproductab}  \eeq
where $\ba\cdot \bb :=\frac{1}{2}(\ba \bb + \bb \ba)\in \R$ is the symmetric {\it inner product}, and
$\ba \w \bb := \frac{1}{2}(\ba \bb-\bb \ba)$ is the antisymmetric {\it outer product} of the vectors
$\ba$ and $\bb$, respectively. The outer product satisfies $\ba\w \bb = i(\ba \times \bb)$, expressing
the duality relationship between the standard Gibbs-Heaviside cross product $\ba \times \bb$ and the
outer product $\ba \w \bb \in \G_3^2$. A great advantage of the geometric algebra $\G_3$ over the
Gibbs-Heaviside vector algebra is the cancellation property
\[  \ba \bb = \ba \bc \iff \ba^2\bb = \ba^2\bc \iff \bb = \bc, \]
provided $\ba^2=|\ba|^2 \ne 0$. The equation $\ba \bb = \ba \bc$ forces equality of {\it both}
the scalar and bivector parts of (\ref{gaproductab}).

The triple products $\ba \cdot (\bb \w \bc)$ and $\ba \w \bb \w \bc$ of three
vectors are also important. Similar to (\ref{gaproductab}), we write
\beq \ba (\bb \w \bc) = \ba \cdot (\bb\w \bc)+\ba \w (\bb \w \bc), \label{gaproductabc} \eeq
where in this case
\[ \ba \cdot (\bb \w \bc):=\frac{1}{2}\big(\ba (\bb \w \bc)-(\bb\w \bc)\ba\big)= - \ba \times(\bb \times \bc)\in \G_3^1, \]
and
\[ \ba \w (\bb \w \bc):=\frac{1}{2}\big(\ba (\bb \w \bc)+(\bb\w \bc)\ba\big)=  \ba \cdot (\bb \times \bc)
 i\in \G_3^3. \]
 We refer the reader back to Figure \ref{picbasis} for a picture of the bivector $\ba\w \bb$ and the
 trivector $\ba \w \bb \w \bc$.
 
A much more detailed treatment of $\G_3$ is given in \cite[Chp.3]{SNF}, and in \cite{S08} I explore the close geometric
relationship that exists between geometric algebras and their matrix counterparts. 
Geometric algebra has been extensively developed by many authors over the last 40 years as 
 a new foundation for much of mathematics and physics. See for example \cite{SNF,H/S,H99,LP97}.

\section{Idempotents and the Riemann sphere}

An {\it idempotent} $s \in \G_3$ has the defining property $s^2=s$. Other than $0$ and $+1$, no other
idempotents exist in the real or complex number systems. As we show below, the most general idempotent in $\G_3$ has 
the form $s=\frac{1}{2}(1+\bm + i\, \bn)$ for $\bm,\bn \in \G_3^1$, where $\bm^2- \bn^2=1$ and $\bm \cdot \bn = 0 $.
 In many respects, idempotents have
similar properties to the eigenvectors of a linear operator. 

Let $g= \alpha+ \bm+ i\, \bn $ be a general non-zero geometric number for
$\bm, \bn \in\G_3^1$ and $\alpha\in \G_3^{0+3}$. In order for $g$ to be an idempotent, we must
have 
\[ g^2 = \alpha^2+\bm^2-\bn^2+2i( \bm\cdot \bn) + 2\alpha(\bm+ i\, \bn) = \alpha + \bm + i\ \bn=g. \]
Equating complex scalar and complex vector parts, gives
\[ \alpha^2+\bm^2+ 2 i( \bm\cdot \bn) -\bn^2 = \alpha, \quad {\rm and}
 \quad 2\alpha (\bm+ i \, \bn)= \bm+ i\, \bn, \] 
from which it follows that 
\[ \alpha = \frac{1}{2}, \quad \bm\cdot \bn = 0, \quad {\rm and} \quad \bm^2-\bn^2 = \frac{1}{4}. \] 
Taking out a factor of $\frac{1}{2}$, we conclude that the most general idempotent $s\in \G_3$ has the form
\beq s=\frac{1}{2}(1+\bm+ i\, \bn), \quad {\rm where} \quad \bm^2-\bn^2 = 1, \ \ {\rm and} \ \ \bm\cdot \bn =0, \label{formidempotent} \eeq
as mentioned previously.

Let us explore the structure of a general idempotent $s=\frac{1}{2}(1+\bm+ i\, \bn)\in \G_3$. Factoring out the
vector $\bm:=|\bm|\hat \bm$, we get
\beq s= \bm \Big(\frac{1}{2}\big(1+\frac{ \hat \bm + i\, \hat \bm \bn}{|\bm|}\big)\Big)  = \bm \hat \bb_+, \label{genidempotent} \eeq
 where $\hat \bb_+ = \frac{1}{2}\big(1+\frac{ \hat \bm + i\, \hat \bm \bn}{|\bm|}\big)$ for the unit vector
 $\hat \bb = \frac{ \hat \bm + i\, \hat \bm \bn}{|\bm|}\in \G_3^1$. 
 With a little more manipulation, we find that
 \beq s= s^2 = (\bm \hat\bb_+)^2  = \bm \hat\bb_+ \bm \hat\bb_+ =\bm^2 \hat\bm \bb_+ \hat \bm  \hat\bb_+= \bm^2 \hat \ba_+\hat \bb_+, \label{genidempotenta}   \eeq
 where $\hat\ba_+ := \hat\bm \bb_+ \hat \bm $. Since 
\beq \hat\bm \hat\bb_+ \hat \bm = (-i\hat\bm) \hat\bb_+ (i\hat \bm), \label{rothatb} \eeq
this means that the parallel component of the vector $\hat \bb$ in the plane of the bivector $i\, \hat \bm$
 is being rotated through $\pi$ radians ($180$ degrees) to obtain the vector $\hat \ba$.

Let us further analyse properties of the idempotent $s=\frac{1}{2}(1+\bm +i\, \bn)$
 given in (\ref{formidempotent}) and (\ref{genidempotent}). Since $\bm^2-\bn^2=1$, we can write
\beq \bm+ i\, \bn=\hat\bm  \cosh \phi  + i\, \hat\bn \sinh \phi   = \hat \bm e^{\phi i\,\hat \bm \hat \bn} =
   e^{-\frac{1}{2}\phi i\,\hat \bm \hat \bn} \hat \bm e^{\frac{1}{2}\phi i\,\hat \bm \hat \bn} , \label{boostm} \eeq
where $\cosh \phi:=|\bm|$, and $ \sinh \phi: = |\bn|$ for some $ 0 \le \phi < \infty$.   
The relation (\ref{boostm}) shows that the complex unit vector $\bm+\,i \bn$ can be
 interpreted as being the {\it Lorentz boost}
of the unit vector $\hat \bm \in \G_3^1$ through the {\it velocity} 
\beq \bv/c = \tanh( \phi i\, \hat \bm \,\hat \bn) =-\hat \bm \times 
\hat \bn \tanh\phi. \label{spinvelocity} \eeq
We call $\bv/c$ the {\it spin velocity} associated with the idempotent $s$. 
The spin velocity $\bv=0$ when $\hat \ba = \hat \bm = \hat \bb$, and
the spin velocity $\bv\to c$ as $\hat \bm \to \hat \bb_\perp$ 
and $\hat \ba \to - \hat \ba$, where $\hat \bb_\perp$ is any unit vector
perpendicular to $\hat \bb$.  

There is a very important property of {\it simple idempotents} of the form 
$\hat\ba_\pm=\frac{1}{2}(1\pm \hat\ba)$, where $\hat\ba\in \G_3^1$.
Let $\hat\bb_+=\frac{1}{2}(1+\hat \bb)$ be a second simple idempotent. Then
\beq \hat \ba_+ \hat\bb_+\hat\ba_+= \frac{1}{2}(1+\hat\ba \cdot \hat\bb)\hat\ba_+.  \label{prodidem} \eeq
This property is easily established with the help of (\ref{gaproductab}), 
\[ \hat \ba_+ \hat\bb_+\hat\ba_+= \frac{1}{4} (1+ \hat \ba)(1+ \hat\bb)\hat \ba_+=\frac{1}{4}(1+\hat \ba+\hat \bb +\hat \ba \hat \bb )\hat \ba_+\]
\[=\frac{1}{4}(1+\hat \ba+\hat \bb -\hat \bb \hat \ba + 2 \hat \ba\cdot \hat \bb)\hat \ba_+= \frac{1}{4}(2 
+ 2\hat \ba\cdot \hat \bb)\hat \ba_+= \frac{1}{2}(1+\hat\ba \cdot \hat\bb)\hat \ba_+. \]
Since $s=\bm^2 \hat\ba_+\hat\bb_+$ in (\ref{genidempotenta}) is an idempotent, it easily follows from (\ref{prodidem}) that
\beq \bm^2=\frac{2}{1+\hat \ba \cdot \hat \bb}. \label{propm2} \eeq
Another consequence of (\ref{prodidem}) that easily follows is
\beq  \hat \ba_+ \hat\bb\hat\ba_+=(\hat \ba \cdot \hat \bb) \hat \ba_+.  \label{conprodidem} \eeq
  
As an example of a general idempotent (\ref{formidempotent}), consider $s=(1+z\, \be_1)u_+$ where
 $z=\frac{\alpha_1}{\alpha_0}=x+iy$ 
for any $\alpha_0, \alpha_1 \in \G_3^{0+3}$, and $\alpha_0 \ne 0$. It follows from (\ref{formidempotent}),
 that 
 \[ \bm =2\langle s \rangle_1 =\langle (1+z\, \be_1)(1+\be_3) \rangle_1= \langle (1+z\, \be_1 -i\, z\, \be_2 +\be_3) \rangle_1 \] 
 \[=  \frac{z+\con z}{2}\,\be_1+\frac{z-\con z}{2i}\,\be_2+\be_3  =x\,\be_1+y\,\be_2+\be_3\in \G_3^1, \] 
 or in terms of $\alpha_0$ and $\alpha_1$,
\beq \bm =\frac{1}{2\alpha_0\con\alpha_0}\Big((\alpha_1 \con\alpha_0+\alpha_0 \con\alpha_1)\be_1
-i(\alpha_0 \con\alpha_1-\alpha_1 \con\alpha_0)\be_2 + 2\alpha_0\con\alpha_0\be_3\Big). \label{defrotm} \eeq

 Noting that $i\, \bn = \bm \w \be_3$, and using (\ref{genidempotenta}), we  now calculate 
\beq s=\frac{1}{2}(1+\bm+ i \, \bn)= \frac{1}{2}(\bm+ \bm \be_3)      = \bm u_+= \bm^2 \hat \ba_+ u_+, \label{canform2spinor} \eeq
\[ \hat \ba = \hat\bm\, \be_3 \hat\bm=\frac{1}{1+z \con z}\Big((z+\con z)\be_1-i(z-\con z)\be_2+(1-z \con z)\be_3 \Big) \]
\beq = \frac{\Big((\con\alpha_0 \alpha_1+\alpha_0 \con \alpha_1)\be_1+ 
i(\con\alpha_1 \alpha_0-\alpha_1 \con \alpha_0)\be_2+
                            (\con\alpha_0 \alpha_0-\alpha_1 \con \alpha_1)\be_3
                            \Big)}{\con\alpha_0 \alpha_0+\alpha_1 \con \alpha_1}  , \label{compareN} \eeq
and
  \beq \bm^2 =1+z \con z=1+ \frac{\alpha_1 \con \alpha_1}{\alpha_0 \con \alpha_0}
  = \frac{\alpha_0 \con \alpha_0+\alpha_1\con \alpha_1}{\alpha_0\con \alpha_0} .\label{magm2} \eeq 

\begin{figure}
\begin{center}
\no\includegraphics[scale=.35]{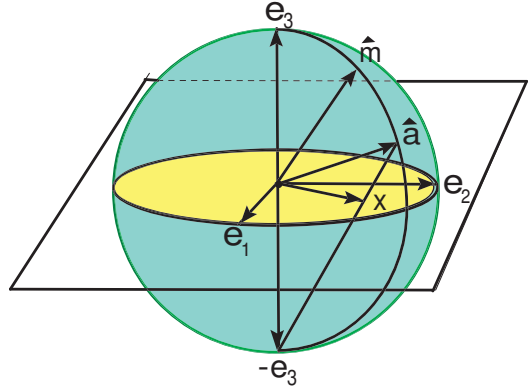}
\caption{Stereographic Projection from the South Pole to the $xy$-plane. }
\label{sterox}
\end{center}
\end{figure}

 The above ideas can be related very simply to the {\it Riemann sphere}. The compact Riemann sphere is
 defined to be the projection of the $xy$-plane onto the the unit sphere $S_2$, centered at the origin, in $\R^3$.
 The stereographic projection from the south pole at the point $-\be_3$, is defined in terms of the projection of $\bm$, given in (\ref{defrotm}), onto
 the $xy$-plane, 
 \[ \bx:=P_{xy} (\bm) =  \frac{z+\con z}{2}\,\be_1+\frac{z-\con z}{2i}\,\be_2=
             x\, \be_1+ y\, \be_2, \]
 that corresponds to the point $\hat \ba \in S_2$ defined in (\ref{compareN}). To check our calculations, we see that
 \beq \bx = t(\hat \ba+\be_3)-\be_3 \quad \iff \quad \bm= \frac{2}{\hat \ba + \be_3}, \label{checkx} \eeq
 for $t=(1+z\con z)/2= \bm^2/2$, so $\bx$ is on the ray passing through the south pole and the point $\hat \ba \in S_2$. Conversely, given
 the point $\hat \ba=a_1 \be_1+a_2 \be_2+a_3 \be_3\in S_2$, we find that
 \[ z = \frac{a_1+i a_3}{1+a_3}=\frac{\alpha_1}{\alpha_0}, \]
 as can be checked using (\ref{compareN}) and (\ref{checkx}). See Figure \ref{sterox}. Stereographic projection is
 just one example of conformal mappings, which have important generalizations to higher dimensions \cite{Sob2012}. 
  
Using (\ref{canform2spinor}), the quantity
   \beq |\alpha\rangle:= \sqrt 2(\alpha_0+\alpha_1 \be_1)u_+ =
   \sqrt 2 \alpha_0 s =\sqrt 2 \alpha_0 \bm u_+ = \sqrt 2 \alpha_0 \bm^2 \hat \ba_+ u_+ \label{spinrep} \eeq 
for $\alpha_0,\alpha_1\in \G_3^{0+3}$ defines what I call a {\it geometric ket-spinor}.
Whereas the spinor $|\alpha \rangle$ is defined for all $\alpha_0,\alpha_1 \in \G_3^{0+3}$,
the idempotent $s$ given in (\ref{canform2spinor}) is only defined when $\alpha_0 \ne 0$. 
However, by a simple trick, even this restriction can be removed, as we will see in the next section.

  \section{Properties of spinors}  
  
  Classically, the {\it Pauli spinor} was introduced by Wolfgang Pauli (1900-1958) to incorporate spin into the
  Schr\"odinger equation for an electron. A Pauli spinor is defined to be a {\it column vector}
  $|\alpha \rangle_p:=\pmatrix{\alpha_0 \cr \alpha_1}$
  in the {\it complex $2$-dimensional Euclidean space} $\C^2$. The notation $|\alpha\rangle_p$,
  called a {\it ket-vector}, is due to Dirac. The corresponding {\it bra-vector} is the
  {\it complex conjugate transpose} 
  \[ \langle \alpha|_p :=\con{|\alpha\rangle}^T_p= \pmatrix{\con\alpha_0 & \con \alpha_1} \]
  of the ket-vector.
 The {\it Euclidean norm} on $\C^2$ is defined by taking the two together to form the {\it bra-ket}
 \beq  ||\alpha\rangle_p|^2=\langle \alpha| \alpha \rangle_p:=
  \pmatrix{\con\alpha_0 & \con \alpha_1}\pmatrix{\alpha_0 \cr \alpha_1}
  =\con\alpha_0 \alpha_0+\con\alpha_1\alpha_1 . \label{normc2} \eeq
 A spinor $|\alpha \rangle_p$ is said to be {\it normalized} if $ \langle \alpha| \alpha \rangle_p=1$.
  
   The complex $2$-dimensional space, with the norm as defined above, behaves like a $4$-dimensional
  real Euclidean space. More generally, for distinct ket-vectors $|\alpha\rangle_p$ and $|\beta\rangle_p$,
  the {\it sesquilinear inner product} is defined by
  \beq  \langle \alpha | \beta \rangle_p:=  \pmatrix{\con\alpha_0 & \con \alpha_1}\pmatrix{\beta_0 \cr \beta_1} 
  = \con\alpha_0 \beta_0+\con\alpha_1\beta_1 \in \C. \label{sesquilinear} \eeq
  
  We replace the Pauli spinor with a corresponding element of a {\it minimal left ideal}, and the
  geometric number (\ref{specbasis3}), (\ref{matrixrep}) that it represents 
  \beq \pmatrix{\alpha_0 \cr \alpha_1}     \quad \longleftrightarrow \quad 
  \pmatrix{\alpha_0 &0\cr \alpha_1 &0}\quad \longleftrightarrow \quad  (\alpha_0+\alpha_1 \be_1)u_+.
  \label{replacespinor}  \eeq
  We then define the spinor $|\alpha\rangle:=\sqrt 2 (\alpha_0+\alpha_1 \be_1)u_+$, using the same
  Dirac ket-notation but without the subscript ``p". 
  The bra-vector is then introduced as the geometric algebra {\it reverse} of
  the ket-vector,
  \[ \langle \alpha |:=|\alpha\rangle^\dagger =\sqrt 2 u_+(\con \alpha_0+\con \alpha_1 \be_1). \] 
  
  For the bra-ket inner product of the ket-spinors $|\alpha \rangle$ and $|\beta \rangle$,
   we first form the geometric product
  \[  \langle \alpha ||\beta\rangle=2(\con \alpha_0+\con \alpha_1 \be_1)
   (\beta_0+\beta_1 \be_1)u_+=2(\con\alpha_0 \beta_0+\con\alpha_1\beta_1)u_+,\]
   and then take the scalar and $3$-vector parts
   \beq  \langle \alpha |\beta\rangle:=\Big\langle \langle \alpha ||\beta\rangle\Big\rangle_{0+3}
   =(\con\alpha_0 \beta_0+\con\alpha_1\beta_1). \label{geoinnerprod} \eeq  
 The extra factor of $\sqrt 2$ was introduced into the definition of a ket-spinor in order to
 to eliminate the unwanted factor of 2 that would otherwise occur
  in the definition of the inner product (\ref{geoinnerprod}).
  
  We see that the spinor inner products (\ref{sesquilinear}) and (\ref{geoinnerprod})
  agree with each other, but whereas the Pauli ket-vector is a complex $2$-component
   column matrix, the corresponding object in $\G_3$ is the geometric ket-spinor given in
   (\ref{replacespinor}). Spinor spaces of a left ideal of a matrix algebra were first considered in the $1930$'s by
   G. Juvet and F. Sauter \cite[p.148]{LP97}. The advantage enjoyed by the geometric ket-spinor 
   over the Pauli ket-vector is that the former inherits the unique algebraic properties
   of a geometric number, in addition to a comprehensive geometric significance. It is
   often remarked that the complex numbers play a special role in mathematics and physics because of
   their many almost {\it magical} properties \cite[p.67]{Pen04}. Geometric algebra takes
   away some of the magic by providing a comprehensive geometric interpretation to
   the quantities involved.  
     
   A spinor $|\alpha \rangle$ with {\it norm} $\rho :=\sqrt{\langle \alpha|\alpha\rangle}$
  is said to be {\it normalized} if 
  \[ \rho^2 =\langle \alpha | \alpha \rangle = \alpha_0 \con \alpha_0+\alpha_1 \con \alpha_1 = 1,\]
  from which it follows, using (\ref{magm2}) and (\ref{spinrep}), that any non-zero spinor $|\alpha \rangle$ can be written in the
  perspicuous cannoical form
  \beq |\alpha \rangle = \sqrt 2 \rho e^{i \theta} \hat\bm u_+ \quad \longleftrightarrow \quad \langle\alpha | = \sqrt 2 \rho e^{-i \theta}u_+ \hat\bm,
  \label{perspicform} \eeq
  where $e^{i\theta}:= \frac{\alpha_0}{\sqrt{\alpha_0 \con \alpha_0}}$. 
  Many non-trivial properties of spinors can be easily derived from this form. As a starter,
  from (\ref{prodidem}) and (\ref{perspicform}), we calculate the {\it ket-bra} geometric product of
   the normalized geometric ket-spinor $|\alpha\rangle$, getting 
      \beq \frac{1}{2}|\alpha \rangle \langle \alpha |=\hat \bm u_+ u_+ 
      \hat \bm = \hat \bm u_+ \hat \bm = \hat\ba_+. \label{idempotenta}\eeq
      
   From (\ref{perspicform}), other important canonical forms for a non zero spinor $|\alpha \rangle$ are quickly established.
In particular, using the fact that $\be_3 u_+=u_+$, from 
  \beq |\alpha \rangle = \sqrt 2 \rho e^{i\theta} \hat \bm u_+ =  \sqrt{2} \rho e^{i \theta}\hat \bm \be_3 u_+ =  \sqrt{2} \rho \hat \bm \be_3 e^{i\be_3 \theta}u_+
   = \sqrt{2} \rho  e^{i \theta}\hat \ba  \hat \bm u_+, \label{importantcanon} \eeq
it follows that
  \beq   |\alpha \rangle = \sqrt 2 \rho e^{i\theta} \hat \bm u_+  =  \sqrt{2} \rho e^{i (\theta +\hat \bv \phi)} u_+
  =  \sqrt{2} \rho e^{i \hat \bv \phi} e^{i\be_3 \theta}u_+
  = \sqrt 2 \rho e^{i\hat \bc \omega} u_+, \label{newperspicform} \eeq
  where $e^{\i \hat \bv \phi}:=\hat \bm \be_3$ and $e^{i\hat \bc \omega} := e^{\i \hat \bv \phi}  e^{i\be_3 \theta}$.
  Of course, all these new variables 
  \[ \theta, \phi ,\omega \in \R \quad {\rm and} \quad \hat \ba, \hat \bm, \hat \bv, \hat \bc \in \G_3^1,  \]
  have to be related back to the non-zero spinor $|\alpha \rangle =\sqrt 2 (\alpha_0+\alpha_1 \be_1)u_+$, where
  \[ \alpha_0=x_0+i y_0 \in \G_3^{0+3}, \quad {\rm and} \quad \alpha_1=x_1+iy_1 \in \G_3^{0+3},\]
  for $x_0,y_0,x_1,y_1 \in \R$, which we now do.
  
   We first find, by using (\ref{defrotm}) and (\ref{compareN}), that 
 \beq \bm = \frac{x_0 x_1+y_0 y_1}{x_0^2+y_0^2}\be_1+\frac{x_0 y_1-x_1 y_0}{x_0^2+y_0^2}\be_2+ \be_3, \label{minxym} \eeq
 so that
 \beq \hat\bm = \frac{x_0 x_1+y_0 y_1}{\sqrt{x_0^2+y_0^2}}\be_1+\frac{x_0 y_1-x_1 y_0}{\sqrt{x_0^2+y_0^2}}\be_2+
  \sqrt{x_0^2+y_0^2}\be_3 \label{minxy} \eeq
 from which it follows that $\hat \bv = \frac{\hat \bm \times \be_3}{|\hat \bm \times \be_3|}$, where
 \[ \cos \phi = \hat\bm \cdot \be_3 = \sqrt{x_0^2+y_0^2}, \quad {\rm and} \quad \sin \phi =|\hat \bm \times \be_3|=
 \sqrt{1-x_0^2-y_0^2}, \]
 and where $0< \phi < \frac{\pi}{2}$. We also have
  \[ \cos \theta =\frac{ Re\,\alpha_0}{\sqrt{\alpha_0 \con\alpha_0}}= \frac{x_0}{\sqrt{x_0^2+y_0^2}}, \quad {\rm and} 
  \quad \sin \theta = \frac{-i\, Im\,\alpha_0}{\sqrt{\alpha_0 \con\alpha_0}} =
 \frac{y_0}{\sqrt{x_0^2+y_0^2}}, \]
 where $0\le \theta < 2\pi$. We also use (\ref{compareN}) to compute
 \beq \hat \ba = 2(x_0 x_1+ y_0 y_1)\be_1 +2(x_0 y_1- y_1 y_0)\be_2+ (x_0^2 +y_0^2-x_1^2-y_1^2)\be_3 .\label{minahat} \eeq
  
 Using the immediately preceeding results, and that $\hat \bv \be_3=i \big(\frac{(\hat\bm\cdot \be_3)\be_3-\hat \bm}{\sin \phi} \big)$,
  we now calculate 
  \[  e^{i\hat \bc \omega}=e^{i\hat \bv \phi}e^{i\be_3 \theta} =  (\cos\phi + i\hat \bv \sin \phi) (\cos\theta + i\be_3\sin \theta) \]
  \[= \cos\theta \cos\phi + i\big(\hat \bv\cos\theta \sin\phi  + \be_3\cos\phi\sin \theta \big)- \hat \bv\be_3 \sin \phi \sin \theta\]
 \[ = \cos\theta \cos\phi + i\big( \hat \bv\cos\theta \sin\phi + \hat \bm \sin \theta\big), \]
 and finally,
  \beq e^{i\hat \bc \omega}  =e^{i\hat \bv \phi}e^{i\be_3 \theta} = \cos \omega + i \hat \bc \sin\omega \label{expomega} \eeq
  where
  \[  \hat \bc = \frac{\hat \bm \times \be_3 \cos \theta +\hat \bm \sin \theta}{\sqrt{1-\cos^2 \theta \cos^2 \phi}} 
                  =\frac{y_1 \be_1-x_1\be_2+y_0 \be_3}{\sqrt{1-x_0^2}},  \]
  and
  \[ \cos \omega = \hat\bc \cdot \be_3 = x_0, \quad {\rm and} \quad \sin \omega =|\hat \bc \times \be_3|=
 \sqrt{1-x_0^2} \]
 for $0\le \omega < \pi$. All of these straight forward calculations are made easy with Mathematica!    
      
\begin{figure}
\begin{center}
\no\includegraphics[scale=.40]{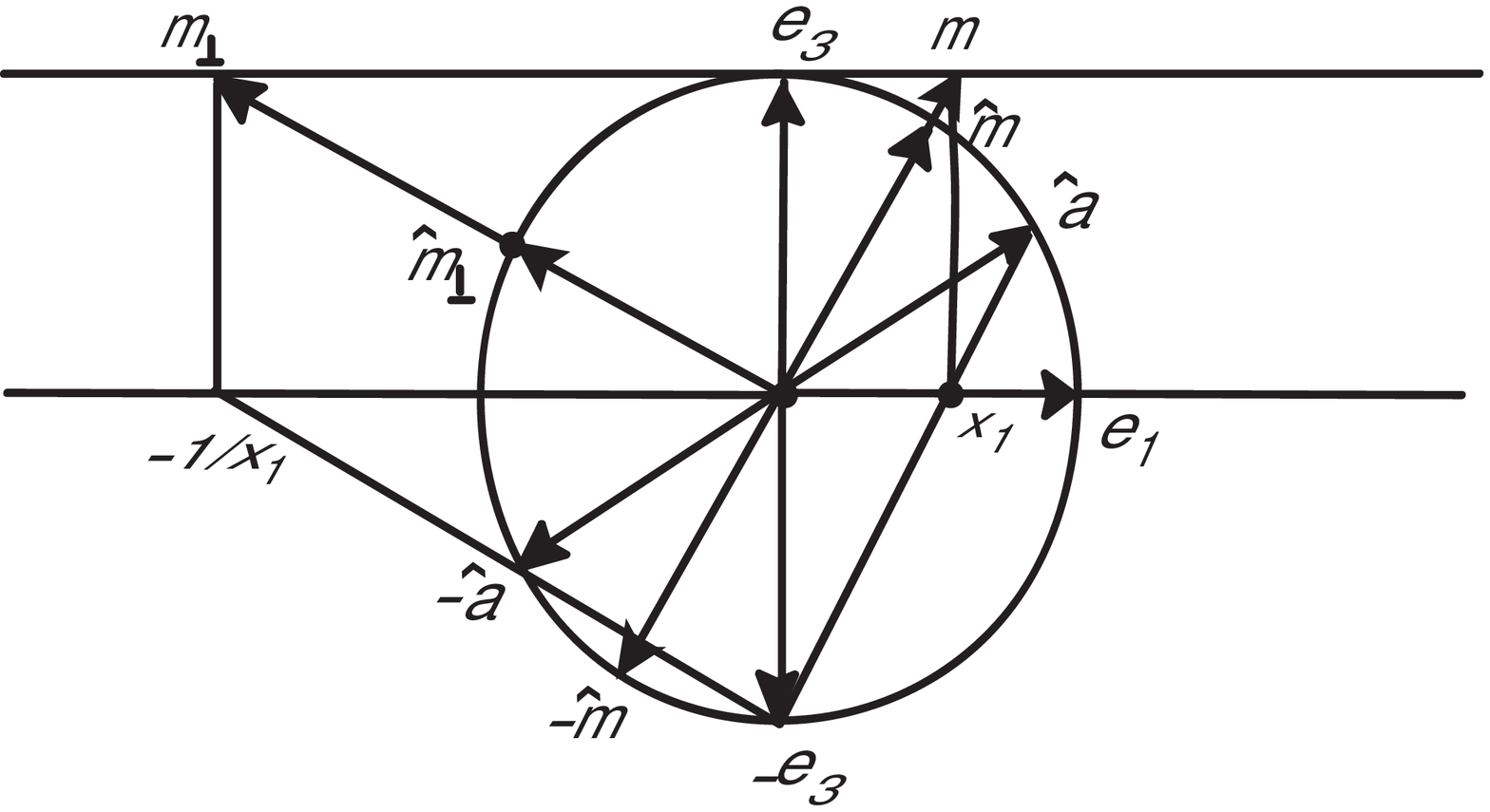}
\caption{Cross section of the unit sphere in the $xz$-plane. }
\label{crosection}
\end{center}
\end{figure}

   Let us take a cross section of the Riemann sphere, shown in Figure \ref{sterox}, to get
   a better understanding of the nature of a spinor. Whereas any cross section through the
   north and south poles would do,
    we choose the great circle obained by taking the intersection of
   the $xz$-plane ($y=0$) with the Riemann sphere, see Figure \ref{crosection}. 
   As the point $\hat \bm$ is moved along this great circle, the
   point $\hat \ba$ moves in such a way that $\hat \bm$ is always at the midpoint of the arc
   joining the points $\be_3$ at the north pole and the point $\hat \ba$. When the point 
   $\hat \bm=\be_3$, the three points coincide.  Choosing $\alpha_0=x_0=1$, and $\alpha_1=x_1$ the geometric ket-spinor
   which determines {\it both} the points $\hat \ba$ and $\hat \bm$
    is $| \alpha\rangle = \sqrt{2} (1+x_1\,\be_1)u_+$,
    corresponding to the Pauli ket-vector $\pmatrix{1 \cr x_1}$.
    Using (\ref{minxy}) and (\ref{minahat}), the unit vectors 
   \[\hat \ba=\frac{1}{1+x_1^2}\Big(2x_1 \, \be_1+(1-x_1^2)\be_3\Big), \quad {\rm and} \quad
   \hat \bm =\frac{1}{1+x_1^2}(x_1\, \be_1 +\be_3),\]
  and $\hat \ba$ is on the ray
   eminating from the South Pole at $-\be_3$. This ray crosses the $x$-axis at the point $x_1 \be_1$, which
   also determines, by (\ref{minxym}), the point $\bm=x_1\, \be_1+\be_3$.

    Let us now rotate the unit vector starting at the point $\hat \bm$, counterclockwise
     past the North Pole at $\be_3$, until
    it reaches the point $\hat\bm_\perp$ perpendicular to the unit vector $\hat \bm$.
    The vectors $\bm=\bx +\be_3 $ and $\bm_\perp$, are specified by the general relationship
\beq \bm_\perp=\frac{1}{\bm\w \be_3}\bm =
\frac{1}{\bx \be_3}\bm =\frac{\be_3 \bx}{\bx^2}(\bx+\be_3)=-\frac{1}{\bx}+\be_3, \label{genperprel} \eeq
so that for $\bm=x_1\be_1+\be_3$,
    \[  \hat \bm =\frac{1}{\sqrt{1+x_1^2}}(x_1\, \be_1+\be_3), \quad \hat \bm_\perp =\frac{1}{\sqrt{1+x_1^2}}(-\be_1+x_1 \be_3).\]
The unit vectors $\hat \bm$ and $\hat \bm_\perp$ are clearly orthogonal since
 $\hat \bm \cdot \hat \bm_\perp=0$. 
 Thus, when the unit vector $\hat \bm$ is turned counterclockwise through $90$ degrees, the unit vector
 $\hat \ba$ is rotated through $180$ degrees into $-\hat \ba$. The corresponding geometric ket-spinor
 to $\hat \bm_\perp$ is $|\beta\rangle=\sqrt 2 (1-\frac{1}{x_1}\be_1)u_+$. The details
 of this construction are shown in Figure \ref{crosection}.
 
 If we continue rotating $\hat \bm$ counterclockwise to a point where it
 crosses the equator into the southern hemisphere, the ket-vector corresponding $\hat \bm$ and
 $\hat \ba$ will be $-\pmatrix{1 & x_1}^T$, in accordance with the fact that
 \[ (-\hat \bm)\be_3(-\hat \bm)=\hat \bm  \be_3 \hat \bm.\]
 Continuing rotating $\hat \bm$ counterclockwise returns $\hat \ba$ to its original position, 
 but the Pauli ket-vector
 representing it will now be $-\pmatrix{1 & x_1}^T$. Because the rotation (\ref{compareN})
is tethered to the point $\be_3$, it is only after rotating $\hat \bm$ through another 180 degrees
that it will return to its original position. This {\it double covering} behaviour will only
occur when $\hat \bm$ is rotated on great circles passing through the North and South poles. For example,
when $\hat \bm$ is rotated on the circle at a latitude of $45$-degrees North, the point $\hat \ba$ will
follow an orbit around the equator directly beneath $\hat \bm$, and hence one complete rotation of
$\hat \bm$ will product one complete rotation of $\hat \ba$, bringing its ket-vector back to
its original value without attaching a minus sign.

 
 In quantum mechanics, a normalized spinor represents the
 {\it state} of a particle which evolves in time by a unitary transformation. 
 This should not be surprising in so far as that by (\ref{newperspicform}), any normalised spinor 
 $|\alpha \rangle$ can be expressed in the cannocal form
 \[|\alpha\rangle = \sqrt 2 (\alpha_0+\alpha_1 \be_1)u_+ =\sqrt 2 e^{i\hat \bc \omega}u_+, \] 
 where $e^{i\hat \bc \omega}$ is a {\it unitary transformation}, since $ e^{i\hat \bc \omega}\big(e^{i\hat \bc \omega}\big)^\dagger
 =e^{i\hat \bc \omega}e^{-i\hat \bc \omega}=1$.
   I will return
 to these ideas shortly, after I discuss {\it Cartan spinors} and {\it spinor operators}.
 
 \section{Cartan spinors and spinor operators} 
 
 Elie Cartan defines a {\it $2$-component spinor} $\pmatrix{\alpha_0 \cr \alpha_1}$ to represent a null complex vector 
 \[ N=\br + i\, \bs=\sum_{k=1}^3 (r_k+i\,s_k)\be_k= \sum_{k=1}^3z_k \be_k,\ \ {\rm where} \ \
 z_1^2+z_2^2+z_3^2=0,\]
 for $z_k \in \G_3^{0+3}$,
  \cite[p.41]{C81}. We find that 
  \[ N=\Big(\sqrt 2(\alpha_0+\alpha_1 \be_1)u_+\Big) \be_1 \Big(\sqrt 2(\alpha_0+\alpha_1 \be_1)u_+\Big)^*
   \]
   \beq =2(\alpha_0+\alpha_1 \be_1)u_+ (\alpha_0-\alpha_1 \be_1)\be_1 = (\alpha_0^2-\alpha_1^2)\be_1+(\alpha_0^2+\alpha_1^2)i\, 
   \be_2 - 2\alpha_0 \alpha_1 \be_3.  \label{Nforma} \eeq  
 Following Cartan, we solve these equations for $\alpha_0, \alpha_1$ in terms of $z_k$ for $k=1,2,3$, to get
 \[  \alpha_0 = \pm \sqrt{\frac{z_1- i z_2}{2}} \quad {\rm and}
  \quad \alpha_1 = \pm i\sqrt{\frac{ z_1+ i z_2}{2}}. \]
   
  Using (\ref{perspicform}) and (\ref{Nforma}), 
  we can solve for $N$ directly in terms of the idempotents $\hat \ba_\pm$ and the vector $\hat\bm \,\be_1 \hat\bm$.
 We find that
 \beq N =-2 \rho^2 e^{2i \theta} \hat \ba_+ \hat\bm\, \be_1 \hat\bm 
  =-2 \rho^2 e^{2i\theta}\hat \bm\, \be_1 \hat\bm \,\hat \ba_-.  \label{Nformfin} \eeq
  We can also represent the null complex vector $N$ explicitly in terms of the unit vector $\hat\bm$ and
   the null vector $\be_1 u_-=\be_1+i \be_2$, or in terms of $\hat \ba_+$ and a reflection of the vector $e^{\be_{12}\theta}\be_1$ in the plane of $i\,\hat \bm$, 
   \beq N =- \rho^2 e^{2i\theta} \hat\bm (\be_1+ i\, \be_2)\hat\bm =-2 \rho^2 \hat \ba_+ \hat \bm\, e^{2 \be_{12}\theta}\be_1\hat \bm . \label{N2ndform} \eeq 
    Whereas $\hat \ba$, in (\ref{compareN}), is a {\it rotation} of the unit vector $\be_3$ in the plane of $i\,\hat\bm$, the null
  vector $N$ is a {\it reflection} of the complex null vector $\be_1 + i \, \be_2$
   in the plane $i\,\hat\bm$.
   
   The {\it spinor operator} of a ket-spinor $|\alpha \rangle=\sqrt{2} (\alpha_0+ \alpha_1 \be_1)u_+$
 is defined by
 \beq \psi  := \frac{1}{\sqrt 2}\big( |\alpha \rangle + |\alpha \rangle^- \big) = (\alpha_0+ 
 \alpha_1 \be_1)u_++ (\con\alpha_0- \con\alpha_1 \be_1)u_- \eeq
 or by using the canonical form (\ref{newperspicform}),
 \beq \psi =\frac{1}{\sqrt 2}\big( |\alpha \rangle + |\alpha \rangle^- \big) = \rho e^{i\hat \bc \omega}u_+ +\rho e^{i\hat \bc \omega}u_- =\rho e^{i\hat \bc \omega}.
  \label{spinops} \eeq
 The matrix $[\psi]$ of the spinor operator $\psi$ in the spectral basis (\ref{specbasis3}) is given by
 \[  [\psi]=\pmatrix{\alpha_0 & -\con \alpha_1 \cr \alpha_1 & \con \alpha_0} = \rho [e^{i \hat \bc \omega}]. \]
 
 If $|\alpha \rangle$ is a normalized spinor, then its spinor operator satisfies
 \[ \det [\psi] =[e^{i \hat \bc \omega}]= \det \pmatrix{\alpha_0 & -\con \alpha_1 \cr \alpha_1 & \con \alpha_0} = \alpha_0 \con \alpha_0+\alpha_1 \con \alpha_1 = 1 ,\]
 and
  \[ [\psi \psi^\dagger]= [\psi][\psi]^* = \pmatrix{1 & 0 \cr 0 & 1}. \]
It follows that $\psi=e^{i \hat \bc \omega}$ defines a unitary transformation in the group $SU(2)$. 
Note also, by using (\ref{newperspicform}), it is easy to show that
 \beq \psi \be_3 \psi^\dagger = \hat \bm \be_3 \hat \bm = \hat \ba,\ \ {\rm and} \ \  \psi u_+ \psi^\dagger = \hat \bm u_+ \hat \bm = \hat \ba_+. \label{spinop2} \eeq 
 The close relationship (\ref{spinops}) and (\ref{spinop2}) between geometric ket-spinors and spinor operators, has been used by D. Hestenes to
 develop an elaborate theory of the electron in terms of the later \cite{H10}. See also \cite{SMex2015}.

\section{The magic of quantum mechanics}

Quantum mechanics, even more than Einstein's great theories of relativity, have transformed our
understanding of the microuniverse and made possible the marvelous technology that we all depend
upon in our daily lives. More and more scientists are realizing that our revolutionary understanding
of the microscopic World must somehow be united with the great theories of cosmology based upon
relativity theory. Whereas we will not discuss the famous paradoxes of quantum mechanics, we
can concisely write down some of the basic rules upon which the foundation of this
 great edifice rests.  
 
A geometric number $S\in \G_3$ is an {\it observable} if $S^\dagger = S$. From (\ref{conjuofrev}),
we see that $S=s_0+\bs$ where $s_0\in \R$ is a real number and $\bs=\sum_{k=1}^3s_k\be_k$ is a vector in $\G_3^1\equiv \R^3$. Generally, in 
quantum mechanics an observer is identified with an Hermitian operator. Indeed, using (\ref{paulimatrices1}) and (\ref{coorformg}), we find that
\beq S = \pmatrix{1 & \be_1}u_+[s_0+ \bs]\pmatrix{1\cr \be_1} =  \pmatrix{1 & \be_1}u_+[S]  
\pmatrix{1\cr \be_1} \label{specbasisSS} \eeq
where $[S]=\pmatrix{s_0+s_3 & s_1-is_2 \cr s_1+is_2 & s_0-s_3}$ 
is an Hermitian matrix.

We can immediately write down the eigenvalues and eigenvectors of the matrix $[S]$ by appealing
to properties of the corresponding geometric number $S$. We first express $S$ in terms of its
natural spectral basis $\hat\bs_\pm =\frac{1}{2}(1 \pm \hat \bs)$, getting
\beq S=S(\hat \bs_++ \hat \bs_-) = (s_0 +|\bs|)\hat \bs_+ + (s_0 -|\bs|)\hat \bs_-, \label{specbasisS} \eeq
where $\bs=|\bs|\hat \bs$ for $|\bs|=\sqrt{s_1^2+s_2^2+s_3^2}$. 
The {\it eigenvalues} of the geometric number $S$, and its matrix $[S]$, are $s_0\pm |\bs|$,
and we say that $\hat \bs_\pm$ are its corresponding {\it eigenpotents}, respectively. The standard eigenvectors of the matrix $[S]$
can be retrived by taking any non-zero columns of the matrices $[\hat \bs_\pm]$, respectively. See \cite{SNF,S2,S0,S3} for
more details regarding the construction and use of the spectral basis of a linear operator.

Having found the eigenvalues and eigenpotents of the observable $S$, we can immediately write down the
corresponding {\it eigenvalues} and {\it eigenspinors} of $S$, simply by multiplying the equation (\ref{specbasisS}) on the
right by $u_+$, getting
\[  Su_+ = S(\hat \bs_+u_++ \hat \bs_-u+) = (s_0 +|\bs|)\hat \bs_+u_+ + (s_0 -|\bs|)\hat \bs_-u_+ .\]
Expressed as a spinor equation, this says that
\[ |S\rangle =(s_0 +|\bs|) |\hat\bs_+\rangle + (s_0 -|\bs|)|\hat\bs_- \rangle, \]
where 
\[ |S\rangle:= \sqrt 2 S u_+, \quad    |\hat\bs_+\rangle:= \sqrt 2 \hat \bs_+ u_+, \quad  |\hat\bs_-\rangle:= \sqrt 2 \hat \bs_- u_+. \]
With the help of (\ref{specbasisS}), we now can write down the eigenspinors of the observable $S$
\beq S |\hat \bs_+\rangle = (s_0 +|\bs|) |\hat\bs_+\rangle , \quad  S |\hat \bs_-\rangle = (s_0 -|\bs|) |\hat\bs_+\rangle .\label{eigenspinors} \eeq

The {\it expected value} of an observable $S=s_0 + \bs$ measured in the state $|\alpha \rangle=\sqrt 2\alpha_0\bm u_+$ is defined by
\[  \langle S \rangle:= \langle \alpha |S|  \alpha \rangle = s_0+\bs \cdot \hat\ba, \]
where $\hat \ba = \hat \bm \be_3 \hat \bm$ as before. The expectated value of the observable $S$ is used to calculate
the {\it standard deviation}
\[ \sigma_S^2 := \langle \alpha |\big( S-\langle S \rangle \big)^2|\alpha \rangle =
               |\bs|^2  \langle \alpha |\big( \hat\bs - \hat\bs \cdot \hat \ba \big)^2|\alpha \rangle=(\bs\times \hat \ba)^2\]
 of the expected value of the observable measured in that state.              
               
  We shall now state and prove a special case of Heisenburg's famous {\it uncertainty principle} as a simple vector analysis identity 
  relating areas in $\R^3$.

\begin{uncertainty} \label{uncertainppl} Given the spinor state $|\alpha\rangle=\sqrt 2 \bm^2 \hat \ba_+u_+$, and the two observables 
$ S=s_0+\bs$, and $T=t_0+\bt$, the standard deviations $\sigma_S$ and $\sigma_T$ of measuring 
 the expected values of these observables in the state $|\alpha\rangle$ satisfy the vector identity
\[ (\bs\times \hat \ba)^2(\bt \times \hat \ba)^2= |(\bs \times \hat \ba)\cdot (\bt \times \hat \ba)|^2 + |(\bs \times \bt)\cdot \hat\ba|^2, \]
which directly implies that
\[ \sigma_S^2\sigma_T^2 \ge \langle \bs \times \bt\rangle^2, \]
since the expected value $<\bs \times \bt>=(\bs \times \bt)\cdot \hat \ba$.
\end{uncertainty}   
\no {\bf Proof:} We include the proof to show off the power of the geometric algebra $\G_3$ over the standard
Gibbs-Heaviside vector algebra, at the same time noting that the former fully encompases the later.

\[ (\bs\times \hat \ba)^2(\bt \times \hat \ba)^2=(\bs\times \hat \ba)\Big((\bs\times \hat \ba)
(\bt \times \hat \ba)\Big)(\bt \times \hat \ba) \]
\[=(\bs\times \hat \ba)\Big((\bs\times \hat \ba)\cdot
(\bt \times \hat \ba)+(\bs\times \hat \ba)\w
(\bt \times \hat \ba)\Big)(\bt \times \hat \ba)     \]
\[ =\Big((\bs\times \hat \ba)\cdot
(\bt \times \hat \ba)+(\bs\times \hat \ba)\w
(\bt \times \hat \ba)\Big)\Big((\bs\times \hat \ba)\cdot
(\bt \times \hat \ba)-(\bs\times \hat \ba)\w
(\bt \times \hat \ba)\Big) \]
\[ =\Big((\bs\times \hat \ba)\cdot
(\bt \times \hat \ba)\Big)^2-\Big((\bs\times \hat \ba)\w
(\bt \times \hat \ba)\Big)^2  =\Big((\bs\times \hat \ba)\cdot
(\bt \times \hat \ba)\Big)^2+\Big((\bs\times \hat \ba)\times
(\bt \times \hat \ba)\Big)^2\]
\[= \Big((\bs\times \hat \ba)\cdot
(\bt \times \hat \ba)\Big)^2+\Big((\bs\times \bt)\cdot \hat \ba\Big)^2.\]

\hfill $\square$

It is interesting to note in the {\bf Uncertainty Principle \ref{uncertainppl}}, it is not enough that the observables $S$ and $T$ 
commute for $\sigma_S \sigma_T=0$. If $S= s_0+\bs$ and $T=t_0 + \bs$ and $s_0\ne t_0$, $ST=TS$, but
$\sigma_S \sigma_T = |(\bs \times \hat \ba)\cdot (\bs \times \hat \ba)|=0$ only when $\bs \times \hat \ba=0$.
I have never seen this matter discussed in standard textbooks on quantum mechanics, probably because 
they have enough trouble as it is in just establishing the inequality  $\sigma_S \sigma_T\ge |\langle \bs\times \bt\rangle|$ 
for the observables $S$ and $T$, \cite[p.108]{G95}.

Up until now we have only considered $2$-component spinors in the equivalent guise of the elements in the minimal left ideal $\{\G_3 u_+\}$ of $\G_3$,
 generated by the primitive idempotent $u_+\in \G_3$. Much more generally, 
 a {\it spinor state} $\Psi$ is an element of a finite or infinite dimensional Hilbert 
space ${\cal H}$, and an {\it observable} $S$ is represented by a {\it self-adjoint operator} on $\cal H$.\footnote{The general spinor 
$\Psi$ is not to be confused with the spinor operator $\psi$ discussed in Section 5.}
In the finite dimensional case, a self-adjoint operator is just a Hermitian operator. The interested reader may want to delve into the much more profound theory of the
role of self-adjoint operators on a Hilbert space in quantum mechanics, \cite{mitocw,wikiHilbert}.

In order that a normalized state $\Psi\in \cal H$ represents an experimentally measurable quantity of an observable $S$, it must be an eigenvector
of $S$, meaning that $S \Psi = \alpha \Psi$ for some $\alpha \in \C$, and it must satisfy {\it Schr\"odinger's equation}
\beq i\hbar \frac{\partial \Psi}{\partial t}= H \Psi ,\label{schrodinger} \eeq
where the {\it Hamiltonian operator} $H$ is obtained from the classical Hamiltonian equation
\beq H_C = \frac{1}{2}m\bv^2+V = \frac{1}{2m} \bp^2+V \label{classhamilton} \eeq
for the {\it total energy} of a particle with mass $m$, velocity $\bv$, momentum $\bp=m\bv$, and potential energy $V$, by way of
the substitution rule
\[ \bp \to \frac{\hbar}{i}\nabla, \quad H_C \to i\hbar \frac{\partial}{\partial t},\]
Making this subsitution into (\ref{classhamilton}) gives the quantum mechanical Hamiltonian operator equation
\beq i\hbar \frac{\partial}{\partial t}=-\frac{\hbar^2}{2m} \nabla^2 + V =H \label{hamoperator} \eeq
where $\hbar$ is Planck's constant, $\nabla = \be_1 \frac{\partial}{\partial x}+ \be_2 \frac{\partial}{\partial y}+ \be_3 \frac{\partial}{\partial z}$ is
the standard gradient operator, and $\nabla^2 = \frac{\partial^2}{\partial x^2}+ \frac{\partial^2}{\partial y^2}+ \frac{\partial^2}{\partial z^2}$
is the Laplace operator. When the operator equation (\ref{hamoperator}) is applied to the wave function $\Psi$, we
get the Schr\"odinger equation (\ref{schrodinger}) for $H=-\frac{\hbar^2}{2m} \nabla^2 + V $.

In general, the Hamiltonian $H=H(\bx,t)$ is a function of both the position $\bx=x \be_1+ y \be_2+z \be_3$ of the particle, and time $t$.
If $\Psi$ sastisfies Shr\"odinger's equation (\ref{schrodinger}), then according to the Born interpretation of the wave function $\Psi$,
the probablity of finding the particle in the infinitesimal volume $|d^3\bx|=dx dy dz$ at the point $\bx$ and time $t$ is
$|d^3\bx||\Psi|= |d^3\bx|\sqrt{\Psi^\dagger \Psi} $. The Hamiltonian $H$ for position is an important example of a self-adjoint operator in the 
infinite dimensional Hilbert space ${\cal L}^2$ of square integrable functions on $\R^3$ which has a {\it continuous spectrum} \cite[p.50,101,106]{G95}.

Let us return to our study of an observable $S=s_0+\bs$, whose possible ket-spinor states are characterised
by points $(\alpha_0,\alpha_1)$ in the finite dimensional Hilbert space $\C^2$. From our equivalent
vantage point, our ket-spinor $|\alpha\rangle = \sqrt 2 (\alpha_0+ \alpha_1 \be_1)u_+$ lives in $\G_3$. One of 
the magical properties of quantum mechanics is that if a measurment is taken of an observerable $S$ in the normalized ket-state,
\[ |\alpha\rangle \in \{\G_3 u_+\} ,\]
whose evolution satisfies the Schr\"odinger equation
 \[ i\hbar \frac{\partial}{\partial t}|\alpha\rangle=\big(-\frac{\hbar^2}{2m} \nabla^2 + V\big)|\alpha\rangle = H|\alpha\rangle ,  \]
the {\it probability} of finding the observable $S$ in the normalized ket-state $|\beta\rangle $
is $|\langle \alpha |\beta \rangle|^2$, and the {\it outcome} of that measurment will leave it in the state $|\beta \rangle$.

Using (\ref{prodidem}), (\ref{idempotenta}), and (\ref{importantcanon}) to calculate $|\langle \alpha |\beta \rangle|^2$, we find that
\[ \langle \alpha | |\beta \rangle =2 e^{i(\theta_\beta -\theta_\alpha)} u_+ \hat \bm_a\, \hat \ba\, \hat \bb\, \hat \bm_b u_+,                   \]
so that
\[ |\langle \alpha |\beta \rangle|^2=\frac{1}{8} \Big\langle\Big(\langle \beta | |\alpha \rangle\Big) \Big( \langle \alpha | |\beta \rangle\Big)\Big\rangle_{0+3}
 =\frac{1}{2}(1+\hat\ba \cdot \hat \bb).  \]
From this we see that the probability of finding the particle in the state $|\beta \rangle=|\alpha \rangle$ is $1$,
whereas the probability of finding the particle in the state $|\beta \rangle = \sqrt 2 e^{i( \theta_\alpha +\pi)} \hat \ba \hat \bm u_+$, 
when $\hat \bb=-\hat \ba$, is $0$.
 
 Note that the expectation values of the observable $S=s_0+\bs$, with respect to the normalized eigen ket-spinors 
 $\frac{|\bs_\pm\rangle}{\sqrt{ \langle\bs_\pm|\bs_\pm\rangle}}      =\sqrt 2 \hat \bs_+ u_+$, 
 are
 \[ \frac{ \langle \bs_\pm | S | \bs_\pm \rangle}{ \langle\bs_\pm|\bs_\pm\rangle} = s_0 \pm |\bs|,  \]
 respectively, as would be expected. 

 The quantum mechanics of $2$-component spinors in $\C^2$, or their equivalent as elements of the minimal ideal $\{\G_3 u_+\}$ in $\G_3$, apply
 to a whole host of problems where {\it spin} $\frac{1}{2}$ particles are involved, as well as many other problems in elementary particle physics.
 Much effort has been devoted to the development of a {\it quantum computer}, built upon the notion of the {\it superposition} of the {\it quantum bits}
 \[  |0 \rangle := \sqrt 2 u_+, \quad {\rm and} \quad |1\rangle := \sqrt 2 \be_1 u_+,\]
 defined by the ket-spinor 
 \[| \alpha \rangle =\sqrt 2 (\alpha_0+\alpha_1 \be_1)u_+ = \alpha_0 |0 \rangle + \alpha_1 |1 \rangle. \] 
 See \cite{HD02} for a discussion of quantum bits using the spinor operator approach. 
 
 There has also recently been much excitement about the quantum phenomenum of {\it neutrino oscillation}, where a neutrino changes its
 {\it flavor} between three different types, {\it electron neutrinos} $\upsilon_\epsilon$, {\it muon neutrinos} $\upsilon_\mu$, and
 {\it tau neutrinos} $\upsilon_\tau$. This variation in flavor implies that neutrinos have {\it mass}, and therefore travel at {\it less}
 than the speed of light, rather than at the speed of light as was previously thought. The observable for a neutrino with three flavors would require a $3\times 3$
 Hermitian matrix, but in the cases of atmospheric neutrinos, or neutrinos propagating in a vacume, it has been found the
 oscillations are largely between only the electron and muon neutrino states, and therefore a $2$ component 
 spinor model is sufficient.

 The general solution to the Schr\"odinger equation   
 \[  i\hbar \frac{\partial |\alpha\rangle}{dt} = H |\alpha \rangle \]
 for a time-independent Hamiltonian $H=s_0 + \bs$ of the form (\ref{specbasisSS}), (\ref{specbasisS}), 
is particularly simple,
 \[ | \alpha \rangle = \sqrt 2 e^{-\frac{iH}{\hbar}t}u_+= \sqrt 2 e^{-i\frac{s_0}{\hbar}t}e^{-i\frac{|\bs|\hat \bs}{\hbar} t}u_+ \]
   \beq  = \sqrt 2 e^{-i\frac{s_0}{\hbar}t}(\cos\frac{|\bs|}{\hbar} t-i \hat \bs \sin \frac{|\bs|}{\hbar} t) u_+
      =\sqrt 2 e^{-i(\frac{s_0}{\hbar}t-\theta)}\hat \bm u_+ , \label{solnschrodinger} \eeq
where $e^{i \theta}$ and $\hat \bm=\frac{\bx + \be_3}{\sqrt{1+\bx^2}}$ are chosen so that $e^{-i\frac{|\bs|\hat \bs}{\hbar} t}u_+=e^{i \theta}\hat \bm u_+$. Solving these relations, we find that
\[ \cos\frac{|\bs|t}{\hbar} = \frac{\cos \theta}{\sqrt{1+\bx^2}}, \  {\rm and} \ 
\hat \bs = -\frac{\bx \sin \theta + \bx \times \be_3 \cos \theta + \be_3 \sin \theta}{\sqrt{\bx^2 + \sin^2\theta} }.\]
If $\theta=0$, the relations simplify considerably, giving
\beq \cos \frac{|\bs|t}{\hbar} = \frac{1}{\sqrt{1+\bx^2}} \quad {\rm and}\quad \hat \bs = \be_3\times \hat \bx .
          \label{neutrinosimple} \eeq
  \begin{figure}
\begin{center}
\no\includegraphics[scale=.30]{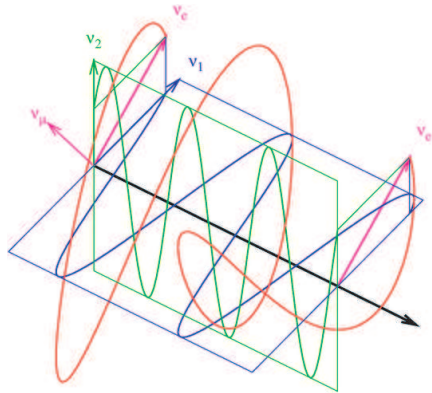}
\caption{Neutrino oscillation between two flavor states, blue and green, and the superposition of
those states, orange.}
\label{fleurotpic}
\end{center}
\end{figure}

Using (\ref{solnschrodinger}),  (\ref{neutrinosimple}), denoting the electron neutrino by the state $|\upsilon_\epsilon\rangle :=|\be_3\rangle$ 
 and the muon neutrino by $|\upsilon_\mu\rangle:=e^{-i\frac{s_0\pi }{2|\bs|}} | \hat \bx \rangle$, then on the great circle of the Riemann sphere with the axis $ \hat \bs$,
  the electron neutrino at the North Pole at time $t=0$, evolves into a muon neutrino at the South Pole at time $t=\frac{\pi \hbar}{2|\bs|}$,
 and then back again, \cite{wiki1,neutrino,neutrino1}. See Figure \ref{fleurotpic}.
  
\section*{Acknowledgement}
I thank Professor Melina Gomez Bock and her students for many lively discussions about quantum mechanics, and
the Universidad de Las Americas-Puebla for many years of support.

\end{document}